\documentclass[aps,pre,twocolumn,superscriptaddress,showpacs,amssymb,a4paper]{revtex4}

\usepackage{amsmath} 
\usepackage{epsfig}

\newcommand{\dt}{\Delta t}
\newcommand{\tw}{t_\mathrm{w}}

\renewcommand{\phi}{\varphi}
\newcommand{\rme}{\mathrm{e}}

\newcommand{\bra}[1]{\langle #1 |}
\newcommand{\ket}[1]{| #1 \rangle}

\newcommand{\nbf}{\boldsymbol{n}}
\newcommand{\qbf}{\boldsymbol{q}}
\newcommand{\rbf}{\boldsymbol{r}}
\newcommand{\ibf}{\boldsymbol{i}}
\newcommand{\jbf}{\boldsymbol{j}}

\newcommand{\oneN}{N^{-1}}

\begin{document}

\title{Activated aging dynamics and negative fluctuation-dissipation
ratios}

\author{Peter Mayer}

\affiliation{Department of Chemistry, Columbia University, 3000
Broadway, New York, NY 10027, USA}

\affiliation{King's College London, Department of Mathematics,
London WC2R 2LS, U.K.}

\author{S\'ebastien L\'eonard}

\affiliation{Laboratoire des Collo\"{\i}des, Verres et Nanomat\'eriaux 
UMR 5587, Universit\'e Montpellier II and CNRS, 
34095 Montpellier Cedex 5, France}

\author{Ludovic Berthier}

\affiliation{Laboratoire des Collo\"{\i}des, Verres et Nanomat\'eriaux 
UMR 5587, Universit\'e Montpellier II and CNRS, 
34095 Montpellier Cedex 5, France}

\author{Juan P. Garrahan}

\affiliation{School of Physics and Astronomy, University of
Nottingham, Nottingham, NG7 2RD, UK}
 
\author{Peter Sollich}

\affiliation{King's College London, Department of Mathematics,
London WC2R 2LS, U.K.}

\date{\today}

\begin{abstract}
In glassy materials aging proceeds at large times via thermal
activation.  We show that this can lead to negative dynamical response
functions and novel and well-defined violations of the
fluctuation-dissipation theorem, in particular, negative
fluctuation-dissipation ratios.  Our analysis is based on detailed
theoretical and numerical results for the activated aging regime of
simple kinetically constrained models. The results are relevant to a
variety of physical situations such as aging in glass-formers,
thermally activated domain growth and granular compaction.
\end{abstract}


\maketitle

Glassy materials display increasingly slow dynamics when approaching
their amorphous state.  At the glass transition, where relaxation times
exceed experimentally accessible timescales, they change from
equilibrated fluids to non-equilibrium amorphous solids.  In the
glassy phase, physical properties are not stationary and the system
ages~\cite{Young}.  A full understanding of the non-equilibrium glassy
state remains a central theoretical challenge.

An important step forward was the mean-field description of aging
dynamics for both structural and spin glasses~\cite{CugKur}.  In this
context, thermal equilibrium is never reached and aging proceeds by
downhill motion in an increasingly flat free energy
landscape~\cite{laloux}.  Two-time correlation and response functions
depend explicitly on both their arguments, and while the
fluctuation-dissipation theorem (FDT) does not hold it can be
generalized using the concept of a fluctuation-dissipation ratio
(FDR). This led in turn to the idea of effective
temperatures~\cite{CugKurPel97}, and a possible thermodynamic
interpretation of aging~\cite{CugKur,FraMezParPel98}.

However, in many systems of physical interest, such as liquids
quenched below the glass transition or domain growth in disordered
magnets, the dynamics is not of mean-field type, displaying both
activated processes and spatial
heterogeneity~\cite{FisHus86,Ediger00}. While some experiments and
simulations~\cite{CriRit03} nonetheless seem to detect a mean-field aging
regime, theoretical studies have found ill-defined
FDRs~\cite{Cri-Vio}, non-monotonic response
functions~\cite{newman,nicodemi,kr,buhot}, observable
dependence~\cite{FieSol02,MayBerGarSol}, non-trivial FDRs without
thermodynamic transitions~\cite{BuhGar02} and a subtle interplay
between growing dynamical correlation lengthscales and FDT
violations~\cite{barrat,kennett}; experiments have also detected
anomalously large FDT violations associated with intermittent
dynamics~\cite{exp}.  It is thus an important task to delineate
when the mean-field concept of an FDR-related effective temperature
remains viable.

Independently of this interpretation, FDRs
have additionally  been recognized as universal amplitude ratios for
non-equilibrium critical dynamics. This makes them important markers for
distinguishing dynamic universality classes. This area has seen
a recent surge of interest~\cite{CalGam}.

Here we study kinetically constrained models, which
are simple models for glassy systems with heterogeneous
dynamics~\cite{RitSol03}. 
We use them to study systematically the impact of activated,
and therefore strongly non mean-field, dynamics on FDRs and
associated effective temperatures. In addition, the dynamics of these
models becomes critical at low temperatures, where dynamical
lengthscales diverge. Our work therefore also pertains
to the study of FDRs in non-equilibrium critical
dynamics. We show that FDT violations retain a simple structure with
well-defined FDRs in the activated regime, elucidate the physical
origin of negative dynamical response functions, and predict the
generic existence of negative FDRs for observables directly coupled to
activated processes.

We focus mainly on the one-spin facilitated model of Fredrickson and
Andersen~\cite{FreAnd84} (FA model), defined in terms of a binary
mobility field, $n_i\in\{0,1\}$, on a cubic lattice; $n_i=1$ indicates that a site is excited (or mobile).  The system evolves through single-site dynamics obeying
detailed balance with respect to the energy function $H=\sum_i n_i$.
The dynamics is subject  
to a kinetic constraint that permits changes at site $i$ only
if at least one nearest neighbor of $i$ is in its excited
state. In any spatial dimension $d$, relaxation times follow an
Arrhenius law at low temperatures~\cite{RitSol03,WhiBerGar04}.

At low temperatures, $T<1$, the dynamics of the FA model is
effectively that of diffusing excitations, which can coalesce and
branch~\cite{WhiBerGar04,SchTri99}. Such a problem can be described in
the continuum limit by a dynamical field theory action with complex fields
$\phi(\rbf,t)$, $\bar{\phi}(\rbf,t)$~\cite{WhiBerGar04,tauber},
\begin{equation}
  S = \int_{\rbf, t} \bar{\phi} (\partial_t - D \nabla^2 - \gamma) \phi -
\gamma \bar{\phi}^2 \phi + \lambda \bar{\phi} (1+\bar{\phi}) \phi^2 ,
\label{equ:action}
\end{equation}
with $n_i \to (1+\bar{\phi}) \phi$ \cite{tauber}.  In terms of the
equilibrium concentration of excitations, $c \equiv \langle n_i
\rangle = 1/(1+e^{1/T})$, the effective rates for diffusion,
coalescence and branching are, respectively, $D \propto c$, $\lambda
\propto c$, and $\gamma \propto c^2$ \cite{SchTri99,WhiBerGar04}.  The
aging dynamics following a quench from $T=\infty$ to low $T$ consists
of two regimes. Initially, clusters of excitations coalesce on timescales
of $O(1)$, reaching a state made of isolated excitations. This
process and the subsequent energy plateau are reminiscent of the
descent to the threshold energy in mean-field models~\cite{CugKur}.
For times larger than $1/D$, excitations diffuse via thermal activation
and decrease the energy as they coalesce.

Following \cite{Lee94} we calculated the connected two-time
correlation to all orders in $\lambda$ at tree
level,
\begin{eqnarray}
N \, C_q (t,\tw) &=& \langle \phi_q(t) \phi_{-q}(\tw) \rangle +
\langle \phi_q(t) \bar{\phi}_{-q}(\tw) \rangle \langle \phi_0(\tw)
\rangle \nonumber \\ &\approx& N \, \tw (\lambda t^2)^{-1} e^{-D q^2
(t-\tw)} f(z).\nonumber
\end{eqnarray}
Here $C_q (t,\tw)$ is the Fourier transform of $\langle n_i(t)
n_j(\tw) \rangle_{\rm c}\equiv \langle n_i(t) n_j(\tw) \rangle -
\langle n_i(t)\rangle \langle  n_j(\tw) \rangle$,
and $\phi_q(t),\bar{\phi}_q(t)$ are those of
$\phi(\rbf,t),\bar{\phi}(\rbf,t)$.  Subscripts $q$ indicate wavevector,
$N$ is the system size, and $z \equiv D q^2 \tw$. We have assumed
that both waiting time $\tw$ and final time $t$ are large ($\gg 1/D$), and set
$\gamma=0$ to get the leading contribution at low $T$.  The function
$f(z)$ goes as $f \approx 1-1/z$ for $z \gg 1$, and $f \approx (1+z)/3$ for
$z \ll 1$.  At $q=0$ we get the energy auto-correlation: $C_0(t,\tw) =
\oneN  \langle H(t) H(\tw) \rangle_{\rm c} 
\approx n(t) \tw /(3t)$, where $n(t) = \langle n_i(t)
\rangle \approx (\lambda t)^{-1}$ is the mean energy per site. These
classical expressions should be accurate above the
critical dimension $d_c$ where fluctuations are negligible.  The limit
$\gamma=0$ corresponds to diffusion limited pair coalescence (DLPC) which has
$d_c=2$ \cite{tauber}.

Consider now a perturbation $\delta H = - h_q A_q$ at time $\tw$, with
$A_q=\sum_i \cos(\qbf \cdot \rbf_i) n_i$.  The action changes by~\cite{note}
$$ T \frac{\delta S}{\delta h_q} = \int_{p,s} \lambda \bar{\phi}_p
\phi_s \phi_{-q-p-s} + \int_p D \bar{\phi}_p \phi_{-q-p} p(p+2q), 
$$
to leading order in $c$. 
For the two-time response function $NR_q(t,\tw) = T
\delta\langle A_q(t)\rangle/\delta h_q(\tw) = 
-T\langle \phi_{-q}(t) \frac{\delta S}{\delta h_q}(\tw) \rangle$ we
then find at tree level
\begin{equation}
  R_q(t,\tw) \approx (\lambda t^2)^{-1} e^{-D q^2 (t-\tw)} (z - 1) .
  \label{Rtree}
\end{equation}
The energy response follows as $R_0(t,\tw) \approx - n(t) / t$.  The
corresponding susceptibility, $\chi_0(t,\tw) = \int_{\tw}^t dt'
R_0(t,t')$ is always negative, and proportional to the density of
excitations at time $t$, $\chi_0(t,\tw) \approx -n(t) (1-\tw/t)$.

The FDR $X_q(t,\tw)$, defined through $R_q(t,\tw) = X_q(t,\tw) \,
\partial_{\tw} C_q(t,\tw)$ if $T$ is included in $R_q$, reads
\[
  X_q(t,\tw) \approx \frac{z-1}{(1+\partial_z) zf(z)}
 \approx \left\{
  \begin{array}{cc} 
    1 - 1/z & (z \gg 1) \\ 
    -3 + 12 z & (z \ll 1) \\
  \end{array} 
  \right. .
\]
At any waiting time $\tw$, for inverse wavevectors $q^{-1}$ 
smaller than the dynamical correlation length, $\xi(\tw) = \sqrt{D\tw}$, FDT is
recovered: $X_q \approx 1$.  However, at small wavevectors, $q \xi\ll
1$, the FDR becomes negative. In the $q \to 0$ limit, one gets the FDR
for energy fluctuations,
\begin{equation}
  X_\infty \equiv X_0(t,\tw) = -3 \quad (d>d_c).
  \label{Xhighd}
\end{equation}
%
%FDT violations thus take a particularly simple form reminiscent
%of the one found in non-equilibrium critical
%dynamics~\cite{MayBerGarSol}. 
This simple form means that on large, non-equilibrated lengthscales
a quasi-FDT holds with $T$ replaced by $T/X_\infty$.  Contrary to pure
ferromagnets at criticality, however, $X_\infty$ is 
{\em negative}. This feature is not predicted by earlier mean-field
studies as it is a direct consequence of thermal activation: if
temperature is perturbed upwards during aging, the dynamics is
accelerated; the energy decay is then faster, giving a negative energy
response.
\begin{figure}
  \begin{center} \includegraphics[width=8.cm]{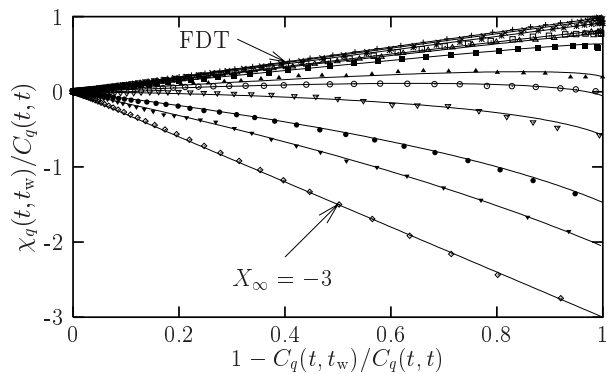}
    \caption{Normalized FD-plots for the FA model at $T=0.1$ in $d=3$
    with fixed $t=2\times10^5$ and running $\tw \in [0,t]$.  Symbols
    are results from simulations, full lines are the tree-level
    predictions.  Wavevectors shown are $q=\pi/x$ with $x=1$, 2, 2.4,
    3, 3.2, 4, 5.33, 6, 8, 12, and 16 (top to bottom).  System size
    was $N=32^3$.  Averages are over $4\times 10^4$ initial
    conditions.}
\end{center}
\end{figure}

The fluctuation-dissipation (FD) plots in Fig.~1 show that the
tree-level calculations compare very well with numerical simulations
of the FA model in $d=3$. We have also confirmed the
diffusive scaling with $z=Dq^2\tw$.  Simulations were performed using
a continuous time Monte Carlo algorithm.  We measure $C_q(t,\tw)$ as the
auto-correlation of the observable $A_q$, on a cubic lattice
with periodic boundary conditions, $N=L^3$ and a linear size $L \gg
\sqrt{D\tw}$. The susceptibility $\chi_q(t,\tw)$ is obtained
by direct generalization of the ``no-field" method of \cite{chatelain}
to continuous time.  We show data using the prescription of
Ref.~\cite{mayer0}, plotting $\chi_q(t,\tw)/C_q(t,t)$ as a function of
$1-C_q(t,\tw)/C_q(t,t)$ for fixed observation time, $t$, and varying
waiting time, $0 < \tw \le t$.  The abscissa runs from 0 ($\tw=t$) to
$\approx$$1$ ($\tw \ll t$).  Using $\tw$ as the running
variable ensures that the slope of the plot is the
FDR $X_q(t,\tw)$~\cite{MayBerGarSol}.  Other procedures, e.g.\ keeping
$\tw$ fixed, would give very different 
results~\cite{Cri-Vio,nicodemi,newman,kr,buhot} that
could lead to erroneous conclusions.

In dimensions $d<d_c$ we need to take fluctuations into account.  For
$d=1$ exact scaling results can be derived for the FA model by
considering a DLPC process with
diffusion rate $D = c/2$.  Because the long-time behavior of DLPC
dynamics is diffusion controlled~\cite{tauber} we are free to choose
the coalescence rate as $\lambda = D$ without affecting scaling
results in the long-time regime ($t,\tw\gg 1/c$)~\cite{MaySol05}.
We focus on the low temperature dynamics
($c \to 0$) and times $t\ll 1/\gamma\propto 1/c^2$ where branching
can be neglected; the system is then still far from equilibrium as it ages.

\begin{figure}
  \begin{center} \includegraphics[width=8.cm]{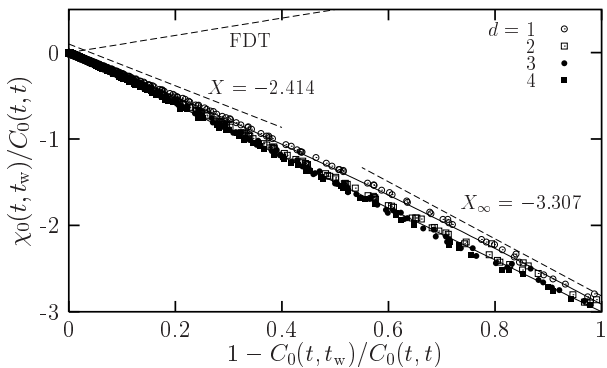}
    \caption{\label{FAenergy} Normalized energy FD-plots for the FA model in
    dimensions $d = 1$ to $4$, for fixed times $t$ and running waiting
    times $\tw \in [0,t]$.  
    Symbols show simulation data, for $T=0.2$ to 0.06 and $t=10^4$ to $10^8$.
    Full curves show
    a  straight line with slope $X_\infty = -3$ 
and the exact limit plot in $d=1$ with 
asymptotic slopes shown as dashed lines.}
\end{center}
\end{figure}

To analyze our DLPC process we use the standard
quantum mechanical formalism~\cite{Kre95}: probabilities are mapped to 
states $\ket{P(t)} = \sum_{\nbf} p(\nbf,t) \ket{\nbf}$, where
$\ket{\nbf} = \ket{n_1,\ldots, n_N}$, so that the master equation reads
$\partial_t \ket{P(t)} = W_C \ket{P(t)}$, with $W_C$ the DLPC
master operator.  Correlations are then $\langle A(t)A(\tw)
\rangle = \bra{1} A \rme^{W_C \dt} A \rme^{W_C \tw}\ket{P(0)}$, where $\bra{1} =
\sum_{\nbf} \bra{\nbf}$ and $\dt\equiv t-\tw$.  DLPC dynamics is closely related to
diffusion limited pair-annihilation (DLPA)~\cite{Kre95} by a
similarity transformation $B$: $W_C = B W_A B^{-1}$
where $W_A$ is the DLPA master operator with diffusion rate $D=c/2$
and annihilation rate $2\lambda=c$. Introducing empty and
parity interval observables, $E_{\ibf} = \prod_{k=i_1}^{i_2-1}(1-n_k)$ and $P_{\ibf} =
\prod_{k=i_1}^{i_2-1}(1-2n_k)$ respectively, where $\ibf=(i_1,i_2)$,
one has $\bra{1}
E_{\ibf} \rme^{W_C \dt} = \bra{1} P_{\ibf} \rme^{W_A \dt} B^{-1}$,
$\bra{1} B = \bra{1}$ and $\bra{1} n_i B = 2\bra{1}
n_i$~\cite{Kre95}. If we interpret the $n_i$ in DLPA as domain walls in an Ising
spin system, $n_i = \frac{1}{2}(1-\sigma_i\sigma_{i+1})$, then
$P_{\ibf} \equiv \sigma_{i_1} \sigma_{i_2}$, and $W_A \equiv W_\sigma$ becomes
the $T=0$ master operator for the Glauber-Ising spin chain.
Substituting the two-spin propagator $\bra{1} \sigma_{i_1}
\sigma_{i_2} \rme^{W_\sigma c \dt}$ derived in \cite{MaySol04} and
mapping back to DLPC then gives
\begin{equation}
  \bra{1} E_{\ibf} \rme^{W_C \dt} = H_{i_2-i_1}(2 c \dt) \bra{1} +
  \sum_{j_1 < j_2} G_{\ibf,\jbf}^{(2)}(c \dt) \bra{1} E_{\jbf},
  \label{equ:prop}
\end{equation}
where $G_{\ibf,\jbf}^{(2)}(\cdot)$ and $H_n(\cdot)$ are given in
\cite{MaySol04}.
%
%As motivated above, the difference between
%Eq.~(\ref{equ:prop}) and the empty interval propagator, $\bra{1}
%E_{\ibf} \rme^{W \dt}$, for the true FA master operator $W$ 
%will be negligible in the low-temperature, long-time,
%far-from-equilibrium aging regime we consider.
%
\begin{figure}
\begin{center} \includegraphics[width=8.cm]{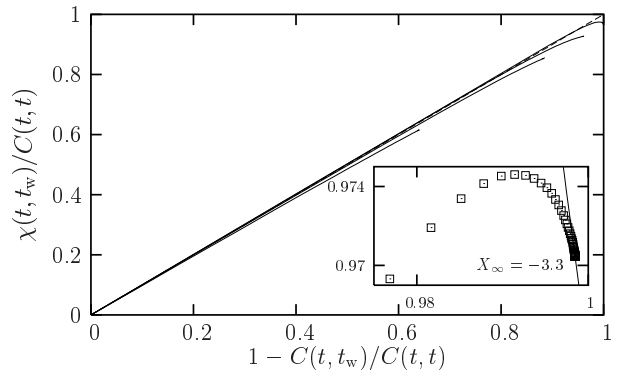}
\caption{\label{FAauto} Normalized FD-plot for the local observable
$n_i$ in the $d=1$ FA model at $T=0.1$ and $t=2\times 10^4$,
$5\times10^4$, $10^5$ and $10^6$ (bottom to top).  The dashed diagonal is
FDT. Inset: final part of the $t=10^6$ data, enlarged to show the
non-monotonic response and $X_\infty\approx -3.307$ (full line).}
\end{center}
\end{figure}
Eq.~(\ref{equ:prop}) together with the identity $n_i=1-E_{(i,i+1)}$
is the key ingredient for the calculation
of the two-time energy correlation and response functions $C_0(t,\tw),
R_0(t,\tw)$ in the $d=1$ FA model~\cite{MaySol05}.  Qualitatively a 
picture similar to $d > d_c$ emerges, but with the FDR now showing a
weak dependence on the ratio $t/\tw$. At equal times,
$X_0(t/\tw \to 1) = -(1+\sqrt{2}) = - 2.414\ldots$, while for 
$t/\tw\to\infty$ the FDR approaches
\begin{equation}
  X_\infty
% = X_0\left(\frac{t}{\tw} \to \infty\right) 
  = {3\pi}/({16-6\pi}) = -3.307\ldots \quad(d=1).
  \label{equ:Xinf}
\end{equation}

Fig.~2 demonstrates complete agreement between our scaling predictions
and simulation data in all dimensions.  In $d = 1$ the data fall on the
curved limit FD-plot~\cite{MaySol05} derived from (\ref{equ:prop}).  In
dimensions $d=3,4$ the simulations are compatible with a constant FDR
$X_\infty = -3$, Eq.~(\ref{Xhighd}).  The data for $d=d_c=2$ suggest
a slightly curved limit plot, possibly due to logarithmic corrections,
but are still compatible with $X_\infty \approx -3$.

Numerically, the no-field method of \cite{chatelain} becomes
unreliable for small $q$. To get precise data
as shown in Fig.~2 we used instead the exact relation
$$
  2 R_0(t,\tw) = (1-2c) \partial_t n(t) + \partial_{\tw} C_0(t,\tw)
  + C_*(t,\tw) . 
$$
for the energy-temperature response of a general class of kinetically
constrained spin models~\cite{MaySol05}. The quantity $C_*(t,\tw)$ is
defined as $\oneN \langle H(t) U(\tw) \rangle_{\rm c}$, with $U=$
$\sum_i f_i (n_i-c)$ and $f_i\in\{0,1\}$ the kinetic constraint.

The auto-correlation $C(t,\tw) = \langle n_i(t)
n_i(\tw)\rangle_{\rm c}$, and associated auto-response, $R(t,\tw)
= T \delta \langle n_i(t) \rangle / 
\delta h_i(\tw)$, to a local perturbation $\delta H = - h_i
n_i$, also have a negative FDR regime.  Previous 
studies~\cite{BuhGar02,buhot} had suggested that the corresponding
FD-plot has an equilibrium form, even during aging.  
A more careful analysis, however, reveals nonequilibrium
contributions (Fig.~3).  Exact long-time predictions for
$C(t,\tw)$ and $R(t,\tw)$, and the resulting FDR, can be derived from
(\ref{equ:prop})~\cite{MaySol05}.  One finds that, as for the energy,
the FDR has the aging scaling $X(t/\tw)$. It crosses over from
quasi-equilibrium behavior,
$X \approx 1$, at $t/\tw\approx 1$, to $X \approx X_\infty$ for $t/\tw
\gg 1$; notably, $X_\infty$ here is the {\em same} as for the energy,
Eq.~(\ref{equ:Xinf}).  However, the region in the FD-plot that reveals the
nonequilibrium behavior shrinks as $O(1/\sqrt{t})$ as $t$ grows.
This explains previous observations of pseudo-equilibrium and makes a
numerical measurement of $X_\infty$ from the local observable $n_i$
very difficult. On the other hand, Fig.~2 shows that with the
coherent counterpart, i.e.\ the energy,  
this would be straightforward, confirming a point made
in~\cite{MayBerGarSol}.  Numerical simulations for $d>1$ produce
local FD-plots analogous to Fig.~3 but with $X_\infty = -3$.  We emphasize
that the non-monotonicity observed is not an artefact of using $t$ as
the curve parameter, as in e.g.~\cite{Cri-Vio,kr}.

\begin{figure}
\begin{center} \includegraphics[width=8.cm]{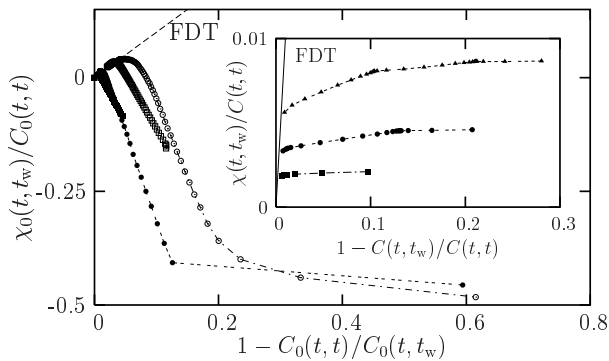}
\caption{\label{East} Normalized energy FD-plots for the East model
for $(T,t)= (.15,100)$, $(.15, 10^5)$, $(.2, 50)$ and $(.2, 5000)$ (from left
to right). Inset: Normalized FD-plot for local observables
for $T=0.15$ and $t=10^2$, $10^5$ and $10^7$ (from bottom to top).}
\end{center}
\end{figure}

We have shown that some important components of the mean-field picture
survive in systems with activated dynamics, in particular the concepts
of time sectors (initial relaxation versus activated aging on long
timescales) with associated well-defined FDRs. However, activation
effects make these FDRs {\em negative}. This effect has not previously
been observed in non-equilibrium critical dynamics, and calls into
question whether effective temperature descriptions are possible for
activated dynamics.
Our arguments show that negative non-equilibrium responses should
occur generically for observables
whose relaxation couples to activation effects; activation need
not be thermal, but can be via macroscopic driving, e.g.\ tapping of
granular materials~\cite{nicodemi,DepSti}. To illustrate how quantitative
effects may vary, we consider briefly the East model~\cite{RitSol03} of
fragile glasses, leaving all details for a separate report.
The behavior is richer due to the hierarchical nature of the
relaxation, which leads to plateaux in the energy decay.  
Properly normalized energy
FD-plots nevertheless have a simple structure, with three
regimes (Fig.~4): for given $t$, equilibrium FDT is obeyed at
small time differences $t-\tw$, indicating
quasi-equilibration within a plateau; a regime with a single
negative FDR follows, coming from the activated relaxation process preceding 
the plateau at $t$; finally, the plot becomes horizontal, corresponding to
all previous relaxation stages which decorrelate the system but do
not contribute to the energy response.  Interestingly, in the FD-plots for
auto-correlation and auto-response (Fig.~4, inset), each relaxation 
stage of the hierarchy is associated with a well-defined 
effective temperature, a structure reminiscent of that found in 
mean-field spin glasses~\cite{CugKur}.

This work was supported by the Austrian Academy of Sciences and EPSRC
grant 00800822 (PM); EPSRC grants GR/R83712/01, GR/S54074/01, and
University of Nottingham grant FEF3024 (JPG).


\begin{thebibliography}{99}
\bibitem{note} The form of the diffusion term arises because
hopping is via creation of a defect at the arrival site. The
hopping rate from site $j$ to $i$ depends only on the field at
site $i$.
\end{thebibliography}
\end{document}